\documentclass
[conference, 10pt, two column,final]{IEEEtran}
\usepackage{amsfonts}
\usepackage{amssymb}
\usepackage{graphicx}
\usepackage{amsmath}
\usepackage{epsfig}
\usepackage{epstopdf}
\usepackage{setspace}
\usepackage{color}
\usepackage{cite}
\usepackage{amsmath}
\usepackage{booktabs}
\usepackage[caption=false, font=normalsize, labelfont=sf, textfont=sf]{subfig}

\renewcommand{\vec}[1]{\ensuremath{{\boldsymbol{#1}}}} 
\newcommand{\vv}[1]{\ensuremath{{\vec{#1}}}} 

\newcommand{\uu}[1] {\ensuremath{\widehat{\boldsymbol{u}}^{(#1)}}}
\newcommand{\uU}[1] {\ensuremath{\widehat{\boldsymbol{U}}^{(#1)}}}
\newcommand{\xc}[1] {\ensuremath{x^{(#1)}}}

\newcommand{\Tr}{\text{Tr}}
\renewcommand{\Re}{\text{Re}}
\renewcommand{\Im}{\text{Im}}
\newcommand{\hh}{\boldsymbol{h} }
\newcommand{\y}{\Delta\hh }
\newcommand{\A}{\boldsymbol{A}}

\newcommand{\M}{\boldsymbol{M}}
\renewcommand{\P}{\boldsymbol{P}}
\newcommand{\Q}{\boldsymbol{Q}}
\newcommand{\T}{\boldsymbol{T}}
\newcommand{\Y}{\boldsymbol{Y}}
\newcommand{\Z}{\boldsymbol{Z}}

\newcommand{\anv} {\ensuremath{\alpha}}
\newcommand{\Tt} {\ensuremath{\tau}}
\newcommand{\av} {\ensuremath{\vec{\alpha}}}
\newcommand{\xv} {\ensuremath{\vec{x}}}
\newcommand{\yv} {\ensuremath{\vec{y}}}

\newcommand{\aG} {\ensuremath{\alpha}}

\newcommand{\gG} {\ensuremath{\gamma}}
\newcommand{\lG} {\ensuremath{\lambda}}

\newcommand{\rG} {\ensuremath{\rho}}
\newcommand{\sG} {\ensuremath{\sigma}}
\newcommand{\tG} {\ensuremath{\tau}}
\newcommand{\zG} {\ensuremath{\zeta}}

\newcommand{\rGp} {\ensuremath{{\rho'}}}

\newcommand{\RR}{\ensuremath{\mathbf{R}}}

\begin{document}
\title{Algebraic Solution for Beamforming in Two-Way Relay Systems with Analog Network Coding}
\author{%
Christopher Thron, Ahsan Aziz,  \emph{Member IEEE}  \\}%
\maketitle
\begin{abstract}
We reduce the problem of optimal beamforming for two-way relay (TWR) systems with perfect channel state infomation (CSI) that use analog network coding (ANC)  to 
a pair of algebraic equations in two variables that can be solved inexpensively using numerical methods. 
The solution has greatly reduced complexity compared to previous exact solutions via
semidefinite programming (SDP). Together with the linearized robust solution described in \cite{Aziz14}, it provides a high-performance, low-complexity robust beamforming solution for 2-way relays.
\end{abstract}
\section{Introduction}
\label{sec:intro}
The analog network coding (ANC) technique has been proven to lead to significantly higher throughput in wireless router scenarios\cite{ANC}. In reference \cite{Rui09}, ANC is evaluated in the context of two-way relay (TWR) systems with two single-antenna source nodes communicating via a multi-antenna relay. In that paper, an optimal beamforming solution was derived that made use of the S-procedure to reduce the beamforming problem to a system of linear matrix inequalities, which can then be solved by semidefinite programming \cite{Boyd}. The paper also noted that the optimal solution could be expressed in terms of four complex design parameters. The current paper provides significant simplifications over that result. We reduce the number of design parameters from four complex parameters to two real parameters. Furthermore, we reduce the problem to an unconstrained minimization problem in two real variables, giving rise to a system of two algebraic equations in two unknowns that can be solved inexpensively to arbitrary accuracy  using conjugate gradient or other numerical methods.   

The optimal solution described above applies to the case where the channel state information (CSI) is known exactly, which is commonly designated as the ``nonrobust'' case. In the ``robust'' case, the CSI is only known to a certain tolerance. It was shown in \cite{Aziz14} that given an exact solution for the nonrobust case,  a low-complexity suboptimal robust solution with very high performance can be found.  Thus our algebraic nonrobust solution can be used as part of a complete  low-complexity solution to the robust beamforming problem for two-way relays with ANC. 

The rest of the paper is organized as follows. In Section \ref{sec:system model},  we present the system model and formulate the problem;
in Section~\ref{sec: ProbDef},  we reduce the problem to a much simpler problem; in Section~\ref{sec:exact} we give algebraic solutions to the simpler problem;
 in Section~\ref{sec:cls} we present simulation results for the nonrobust case; in Section~\ref{sec:robust} we describe the suboptimal robust solution, and present simulation results; and in Section~\ref{sec:final} we summarize our conclusions. 

The notations used in this paper are listed as follows. We define $(\cdot)^{T}$,  $(\cdot)^{H}$,  $\bar{(\cdot)}$  as the transpose, 
Hermitian transpose,  and conjugate operations,  respectively. $\Re(\cdot)$ is the real part and $\Im(\cdot)$ is the imaginary part of a complex variable.
We use $\Tr[\cdot]$  to denote the trace of a matrix.
\section{System Model, and Statement of the Beamforming Optimization Problem}\label{sec:system model}
We consider a two-way relay system similar to the one introduced in \cite{Rui09},  which consists of the relay node $R$ and two terminal nodes $S1$ and $S2$. The relay is equipped with $M$ antennas and the terminal nodes are each equipped with a single antenna.
For terminal node $S_i\, (i=1,2)$, we define $p_i$ as the transmit power level and $\hh _i \in \mathcal{C}^{M\times1 }$ as the complex channel gain from node to relay. We further define $\sigma_i^2$ as the noise variance in the received signal at $S_i\, (i=1,2)$, and $\sigma_R^2\boldsymbol{I}$ as the noise covariance for the received signal at $R$, where all noises are assumed to be circularly symmetric complex Gaussian (CSCG). It was shown in   
 \cite{Rui09}  that for an ANC system in which the terminal nodes exchange information in two consecutive time slots under conditions of channel reciprocity (justified in \cite{Zeng11}), 
transmit power at the relay $R$ is given by 
\begin{align}
\label{eq:txpwr}
G(\A) \equiv \|\A\hh _1\|^2 p_1+\|\A\hh _2\|^2 p_2+
\Tr[\A^H\A]\sigma_{R}^{2}, 
\end{align}
where $\A \in \mathcal{C}^{M\times M }$ is the relay's beamforming matrix.
Reference \cite{Rui09} also shows that if the SINR at node $S_i$ is constrained to be at least $\gamma_i\,(i = 1, 2)$, then assuming perfect knowledge of CSI (which is denoted as the ``nonrobust'' case)  the optimization problem to minimize the relay power can be formulated as follows: find (i=1, 2)
\begin{equation}
\label{modOptProblem0}
\A_* = \textrm{arg} \min_{\A}\left[ G(\A)\right] \qquad
 {\rm s.t.} \quad{f_{i}(\A)}\geq \gamma_i \sigma_i^2, 
\end{equation}
where 
\begin{equation}
\label{eq:f_i_def}
f_{i}(\A) \equiv  |{\hh }_{i}^T\A{\hh }_{k}|^2p_k  -  |\|{\hh }_{i}^T \A\|^2\sigma^2_R \gamma_i, ,~(k \equiv 3-i).
\end{equation} 

We note that the problem in (\ref{modOptProblem0}) is not convex in general,  because the constraints are not convex functions.

\section{Reduction to  real-valued rank 2 problem}\label{sec: ProbDef}
In this section we show how (\ref{modOptProblem0}) can be transformed into a much simpler problem with real coefficients. 

It has been shown previously in \cite{Rui09} that the solution $\A_*$  of (\ref{modOptProblem0}) is of rank 2.  Specifically,   $\A_*$ can be expressed as
\begin{equation}
\A_* = \sum_{i, j=1}^2 (a_{*})_{ij} \bar{\hh}_i \hh_j^{H}  =  [ \bar{\hh}_1,~\bar{\hh}_2] a_* [\hh_1^{H}~;~ \hh_2^H], 
\end{equation}
where $a_*$ is a complex $2 \times 2$ matrix.  
The objective function condition and constraints in (\ref{modOptProblem0}) can be rewritten in terms of the matrix $a_*$. The coefficients which appear in this simplified version of (\ref{modOptProblem0}) will be complex in general; but it is possible to further simplify the expressions so that all coefficients are real as follows. First we define
$$
e^{j\theta} \equiv \frac{\hh_2^H \hh_1}{| \hh_2^H \hh_1 |}; \qquad
t_{\pm} \equiv \frac{ ||\, (\hh_1 / \|\hh_1 \| \pm e^{j\theta}\hh_2 / \| \hh_2\|)\, ||}{  \sqrt{2}}.
$$
 We then choose the following  orthonormal basis $\{\vv{e}_+, \vv{e}_{-}\}$ for the space spanned by $\hh_1$ and $e^{j\theta}\hh_2$:
$$\vv{e}_{\pm} \equiv (\hh_1 / \|\hh_1 \| \pm e^{j\theta} \hh_2 / \| \hh_2 \|) / (\sqrt{2}t_{\pm}).$$
The following may be verified, where $r \equiv t_-/t_+$ (note $r>0$):
\begin{align*}
&\vv{e}_+^H \vv{e}_- = \vv{e}_-^H \vv{e}_+= 0; \\
&\hh_1 = \|\hh_1 \| t_+ (\vv{e}_+ + r\vv{e}_-); \quad
\hh_2 = e^{j \theta}\|\hh_2 \| t_+(\vv{e}_+ - r\vv{e}_-).
\end{align*}
%
Since the vectors $\{\vv{e}_+,  \vv{e}_-\}$ defined above span $\{\hh_1,  \hh_2\}$,  we may alternatively write
\[ \A_* = [\bar{\vv{e}}_+~\bar{\vv{e}}_-] \anv [\vv{e}_+^H~;~\vv{e}_-^H], 
 \]
where $\anv$ is a $2 \times 2$ complex matrix.
Using the following rescaled constants ($i=1, 2; k = 3-i$)
\begin{align*}
q_i &\equiv p_i \|\hh_i\|^2 t_+^2 / \sigma_R^2;
~ c_i \equiv p_k \|\hh_1\|^2 \|\hh_2\|^2 t_+^4/ (\gG_i \sG_i^2);\\
~ d_i &\equiv \|\hh_i\|^2 t_+^2 \sG_R^2 / \sG_i^2;
~\Tt_1 \equiv [1\, ;\, r]; ,~\Tt_2 \equiv [1\, ;\, -r],  
\end{align*}
the optimization problem becomes:
\medskip
\begin{equation}
\label{modOptProblem}
\anv_* = \textrm{arg} \min_{\anv}\left[ g(\anv)\right] \qquad
 {\rm s.t.} \quad{f_{i}(\anv)}\geq 1,~(i=1, 2), 
\end{equation}
where $ (i = 1, 2; k = 3-i)$
\begin{align}
\begin{aligned}\label{eq:power}
g(\anv) &\equiv q_1 \| \anv \Tt_1\| ^2 + q_2\| \anv \Tt_2\|^2 + \Tr[\anv^H \anv];\\
 f_i(\anv) &\equiv c_i|\Tt_i^T\anv\Tt_k|^2 - d_i\|\Tt_i^T\anv\|^2.
\end{aligned}
\end{align}
Note that although $\sG_R^2 g(\anv)$ gives the actual power, for convenience's sake we will refer to $g(\anv)$ as the ``power function''. 

The functions in (\ref{eq:power}) can be compactly expressed as quadratic forms. First we define  $(i = 1, 2; k \equiv 3-i)$
\begin{equation}\label{eq:phidef}
 \Tt_{ii} \equiv \Tt_i \Tt_i^T; \qquad
m  \equiv q_1 \Tt_{11} + q_2 \Tt_{22}  +  I. 
\end{equation}
Next,  for any $2 \times 2$ matrix $\anv$ we define the operations:
\begin{equation}
\begin{aligned}
\av &\equiv [\anv_{11}~\anv_{12}~\anv_{21}~\anv_{22}]^T;\\
\underline{\anv} &\equiv 
\left[ \begin{array}{cc}
\anv & 0 \\
0 & \anv  \end{array} \right]; \qquad
\widetilde{\anv} \equiv 
\left[ \begin{array}{cc}
\anv_{11}I & \anv_{21}I \\
\anv_{12}I & \anv_{22}I  \end{array} \right].
\end{aligned}
\end{equation}
Finally we define
\begin{equation}
\M \equiv \underline{m}; \qquad \T_{ki} \equiv \underline{\tau_{kk}} \widetilde{\tau_{ii}};\qquad  \Q_i \equiv c_i T_{ki} -  d_i \widetilde{\tG_{ii}}, 
\end {equation}
where $\M,  \T_{ki}$,  and $\Q_i$ are all real symmetric $4 \times 4$ matrices.
Using this notation,  we have
\begin{equation}\label{eq:power2a}
g(\anv) \equiv  \av^H \M \av; \qquad
 f_i(\anv)  \equiv  \av^H \Q_i\av.
\end{equation}
Note that all the coefficients in (\ref{eq:power2a}) are \emph{real}. We now show that for any locally-optimal complex feasible solution to (\ref{modOptProblem}) with $g$ and $f_i$ as in (\ref{eq:power2a}),   there also exists a  real feasible solution  that achieves the same power. This implies there always exists a globally optimal \emph{real} feasible solution. 

Let us write $\xv \equiv \text{Re}[\av]$ and $\yv \equiv \text{Im}[\av]$. Then we may consider $g$ and $f_i$ as functions of $\xv, \yv$:
\begin{align}
\begin{aligned}
\label{eq:RealMin2}
g(\xv, \yv) &= \xv^T \M \xv + \yv^T \M \yv;\\
f_i(\xv, \yv) &=   \xv^T \Q_i \xv + \yv^T \Q_i \yv, \quad (i=1, 2).
\end{aligned}
\end{align}
In terms of $M$ and $Q$, the KKT conditions corresponding to the minimization problem 
(\ref{modOptProblem}) are $ (\lG_1,  \lG_2 \ge 0)$
\begin{align}
\label{eq:Mxy1}
 M\xv &= \lG_1 \Q_1 \xv  + \lG_2 \Q_2\xv ; \quad
M \yv = \lG_1 \Q_1 \yv  +   \lG_2 \Q_2 \yv, 
\end{align}
where $\lG_i(1 - f_i(\xv, \yv ))  = 0,~(i=1, 2).$
Consider first solutions of (\ref{eq:Mxy1}) where  $f_{i}(\xv,\yv)= 1,\,i=1,2$.
Using (\ref{eq:Mxy1})  we may verify that the \emph{real} beamforming matrix $\zG_x \xv  + \zG_y \yv $ also satisfies the KKT conditions,  for any real $\zG_x$ and $\zG_y$. Furthermore, if $f_i(\zG_x \xv  + \zG_y \yv,0)=1$ $(i=1,2)$   are satisfied with equality, then it can be shown that  $\zG_x \xv  + \zG_y \yv $ is a real solution which satisfies both constraints and has the same power as the complex optimal solution. The proof that suitable $\zG_x,  \zG_y$ can always be found is accomplished on a case-by-case basis using various geometrical arguments. In the case where only one of the two constraints is satisfied with equality, similar arguments may be used. Details may be found in \cite{Thron14}.

\section{Algebraic solution of the real problem}\label{sec:exact}

Define real $4 \times 4$ orthogonal matrices $\uU{1}, \uU{2}$ as ($\rG \equiv 1+ r^2$)
\begin{align} 
\begin{aligned}
\label{eq:U}
\uU{1} &\equiv  \rG^{-1}\left[ \begin{array}{c c c c}   1 & r & r & r^2 \\ -r & 1 &  r^2 & -r \\ 
r & r^2 & -1 & -r \\ -r^2 & r & -r & 1 \end{array} \right]; \\
\uU{2} &\equiv  \rG^{-1}\left[ \begin{array}{c c c c}   1 & r & r & r^2 \\ r & -1 &  -r^2 & r \\ 
-r & -r^2 & 1 & r \\ -r^2 & r & -r & 1 \end{array} \right]. \qquad
\end{aligned}
\end{align}
The columns of \uU{k} form an orthonormal basis of $\RR^4$ for $k=1,2$, so for any vector $\xv \in \RR^4$, we may write
$ \xv \equiv  \uU{1}\xv^{(1)}  \equiv \uU{2} \xv^{(2)}$. It is possible to show that $(k=1,2)$
\begin{equation*}
f_k(\xv) = \xv^T \Q_k \xv = \rG \left( (c_k \rG -d_k) (\xc{k}_1)^2  - d_k(\xc{k}_2)^2 \right).
\end{equation*}   
Consider first the case where both  constraints $f_k(\xv) \ge 1,~(k=1,2)$ are satisfied with equality. Then we have:
\begin{equation}\label{eq:xck}
\xc{k}_1 = \pm \mu_k(\xc{k}_2),
\end{equation}
where 
\begin{equation*}
\mu_k(x) \equiv \sqrt{a_k  +b_k x^2}~~(k=1,2),
\end{equation*}
and
\begin{equation*}
 a_k \equiv \frac{\rG^{-1}}{c_k \rG -d_k};~~  b_k \equiv \frac{d_k}{c_k \rG -d_k}~~(k=1,2).
\end{equation*}
The other components of 
$ \xv^{(k)}$ are uniquely determined via (\ref{eq:xk_det}),
\begin{figure*}[tp]
\begin{equation}\label{eq:xk_det}
\left[ \begin{array}{c}   \xc{1}_3 \\ \xc{1}_4 \\ \xc{2}_3  \\ \xc{2}_4 \end{array} \right]
 = \left[ \uu{1}_3,~\uu{1}_4,~-\uu{2}_3,~-\uu{2}_4\right]^{-1}
 \left[ -\uu{1}_1,~-\uu{1}_2,~\uu{2}_1,~\uu{2}_2\right]  
\left[ \begin{array}{c}  \xc{1}_1 \\ \xc{1}_2 \\ \xc{2}_1  \\ \xc{2}_2 \end{array} \right],
\end{equation}
\end{figure*}
where $ \uu{k}_j$ denotes the $j$'th column of $\uU{k}$.   
(Note  that the vectors $\{\uu{1}_3,\uu{1}_4,~\uu{2}_3,-\uu{2}_4\}$ are linearly independent when $r \neq 0, \infty$.)
Using the first two rows of this matrix, we may construct a matrix $\P$ such that
$\xv^{(1)}  
\equiv \P
\left[  \xc{1}_1 ,~ \xc{1}_2 ,~ \xc{2}_1 ,~ \xc{2}_2 \right]^T$.
Defining $\Z \equiv \P^T (\uU{1})^T \M \uU{1} \P$, we may then write the power function as:
\begin{align*}
g(\xv) =  [\xc{1}_1,~\xc{1}_2,~\xc{2}_1,~\xc{2}_2 ] \Z  [\xc{1}_1,~\xc{1}_2,~\xc{2}_1,~\xc{2}_2]^T. 
\end{align*}
The solutions we are seeking are the \emph{unconstrained} solutions to
\begin{align}\label{eq:UncSoln}
&(\xc{1}_2, \xc{2}_2) = \textrm{arg} \min_{x,y}~\left[\begin{array}{c}\mu_1(x)\\x\\ \pm \mu_2(y)\\y \end{array}\right]^T    
Z   \left[\begin{array}{c}\mu_1(x)\\x\\ \pm \mu_2( y)\\y\end{array} \right].
\end{align}
Note we have left off one of the $\pm$'s because of symmetry--these other solutions will be the negatives of the solutions to (\ref{eq:UncSoln}).
For each choice of sign in (\ref{eq:UncSoln}), it can be shown there is a unique solution. This can be seen geometrically as follows. The solution $\vv{x}$ considered as a point in $\RR^4$ is determined by the intersection of the 
 nested family of strictly convex sets $\{ g(\vv{x}) \le K\}_{K \in \mathbb{R}^+}$ with a strictly convex set $S$ that is the intersection of two strictly convex components of the constraint sets (one component from each constraint in (\ref{modOptProblem})). 
The set $\{ g(\vv{x}) \le K\}$ is equal to the point $0 \in \mathbb{R}^4$, and since $0 \notin S$ it follows that the minimum of $g(\vv{x})$ on $S$ is positive. Because of the convexity of the sets involved, the smallest value of $K$ for which the intersection  $\{ g(\vv{x}) \le K\} \cap S$ is nonempty produces an intersection consisting of a single point, which is the unique  global minimum of the function $g(x)$ under the given constraints. 

In terms of the solution to (\ref{eq:UncSoln}), 
the optimized beamforming matrix (in vector form) is 
\[ \vv{\aG} =\vv{x} =  \uU{1} \P
\left[
\begin{array}{c}
\mu_1 (\xc{1}_2)\\
\xc{1}_2\\
\pm \mu_2(\xc{2}_2)\\
\xc{2}_2
\end{array}
\right].
 \] 
Using the symbolic algebra software Maxima, we may find explicitly
\begin{align*}
&\vv{\aG} = 
\frac{1}{2} \left[ \begin{array}{cccc}
  1    &   r   &1          & r \\
-r     &  1   &  r         & -1 \\   
1/r   &  1   &  -1/r	&  -1 \\
-1    &  1/r & -1        & 1/r
\end{array}
\right]
 \left[ \begin{array}{c} 
\mu_1(\xc{1}_2) \\ \xc{1}_2 \\  \pm \mu_2(\xc{2}_2) \\ \xc{2}_2
\end{array}
\right].
\end{align*}
There are two local optima, corresponding to the $\pm$ sign in the expression. Maxima can also be used to find the expression (\ref{eq:Zexp}) for $\Z$, which utilizes the definitions
\begin{equation*}
\rG \equiv r^2 + 1; ~\rGp \equiv r^2 - 1.
\end{equation*}
\begin{figure*}[tp]
\begin{align}\label{eq:Zexp}
Z =\frac{1}{4r^2} \left[
\begin{array}{cccc}
\rG ( \rGp^2 q_1 + \rG^2 q_2 + \rG)  &  -2r\rG \rGp q_1      	 		& -\rGp^2( \rG(q_1+q_2)+1)    			& 2r \rGp ( \rG q_2 + 1) \\
-2r \rG \rGp q_1                                 &   \rG (4r^2 q_1 + \rG) 		&  2 r \rGp(\rG q_1 + 1)             			& \rGp^2 \\   
 -\rGp^2( \rG(q_1+q_2)+1)  		      &  2 r \rGp(\rG q_1 + 1)		&  \rG ( \rG^2 q_1 + \rGp^2 q_2 + \rG)	& -2 r \rG \rGp q_2  \\
2r \rGp ( \rG q_2 + 1)			      &   \rGp^2					&  -2 r \rG \rGp q_2						& \rG (4r^2 q_2 + \rG)
\end{array}
\right].
\end{align}
\end{figure*}
We consider now the possibility of optimal solutions for which one constraint is satisfied with equality and the other with strict inequality.  First we  suppose that $f_1(\vv{x})=1$ and $f_2(\vv{x}) > 1$ for the optimal solution $\vv{x}$.  
Since $\vv{x}$ satisfies  $f_1(\vv{x})=1$, we may write $(\xc{1}_1)^2 = a_1  +b_1(\xc{1}_2)^2,$ 
where $\vv{x} =  \uU{1}\vv{x}^{(1)}$.
We may also write the power as 
$g(\vv{x}) =  (\vv{x}^{(1)})^T \Y  \vv{x}^{(1)},$
where
$ \Y \equiv (\uU{1})^T \M  \uU{1}$.  
Since $\M$ is positive definite, then so is $\Y$.  It follows that for $\vv{x}$ to minimize the power subject to the constraint  $f_1(\vv{x})=1$, we must have $\xc{1}_3 = \xc{1}_4 = 0$. We thus have
\begin{equation*}
g(\vv{x}) =  \left[ \begin{array}{c} \mu_1(\xc{1}_2)\\ \xc{1}_2\end{array} \right]^T \left[\begin{array}{cc} y_{11} & y_{12} \\ y_{21} & y_{22} \end{array} \right]  \left[ \begin{array}{c}\mu_1(\xc{1}_2) \\  \xc{1}_2  \end{array} \right],
\end{equation*}
where using Maxima we find
\begin{align*}
y_{11} &= \rG^{-1} \left(q_1 \rG^{\prime 2} + q_2 \rG^2 + \rG\right);~
y_{12}=y_{21} = \rG^{-1} (2q_1 r \rG'); \\  y_{22} &= \rG^{-1} (4q_1 r^2 + \rG).
\end{align*}
This expression is minimized when $\xc{1}_2$ satisfies
\begin{equation*}
 y_{11} b_1 \xc{1}_2  + y_{12} \mu_1(\xc{1}_2) + \frac{y_{12}b_1(\xc{1}_2)^2}{\mu_1(\xc{1}_2)} + y_{22}\xc{1}_2 = 0,
\end{equation*}
which leads to
\begin{align*}
0= &\left( (y_{11} b_1  +  y_{22})^2   - 4 y_{12}^2 b_1 \right)b_1 (\xc{1}_2)^4 \\
&+ \left(( y_{11} b_1  +  y_{22})^2 a_1 -4y_{12}^2a_1 b_1\right) (\xc{1}_2)^2   -   y_{12}^2 a_1 ^2.
\end{align*}
Notice there will always be one positive and one (nonphysical) negative solution for $(\xc{1}_2)^2$.  We should consider both the positive and negative solution for 
$\xc{1}_2$, and we may choose the positive root to obtain $ \xc{1}_1 = \mu_1(\xc{1}_2)$.   Finally, we must check whether or not the second constraint is satisfied:
\begin{align*}
1 <& f_2(\xv) \\
= &\rG \left( (c_2 \rG -d_2) (\xc{1}_1 \uu{1}_1 \cdot \uu{2}_1 + \xc{1}_2  \uu{1}_2 \cdot \uu{2}_1)^2 \right.
\\  &\left. ~~~- d_k(\xc{1}_1 \uu{1}_1 \cdot \uu{2}_2 + \xc{1}_2  \uu{1}_2 \cdot \uu{2}_2)^2 \right),
\end{align*}   
which is equivalent to
\begin{equation*}
 \frac{\rG}{\rG'} < \left( (c_2 \rG -d_2) (\rG' \xc{1}_1 -2r \xc{1}_2  )^2  - d_2( 2r \xc{1}_1 + \rG'  \xc{1}_2  )^2 \right).
\end{equation*}   
For solutions which satisfy the second constraint and not the first, the same equations are used except the indices 1 and 2 are exchanged.

In summary, we have found six candidate solutions to be evaluated: two which satisfy both constraints with equality, and 
 four others that satisfy one constraint with equality, and one with strict inequality. The candidate that has the lowest power will be the true optimal solution.

In practice, the candidate solutions must be computed numerically. Four of the candidates (those for which one of the constraints is not satisfied strictly) are obtained using the quadratic formula. The other two (specified by (\ref{eq:UncSoln}))  can be evaluated to the desired precision using numerical methods such as conjugate gradient. Convergence of the Polak-Ribi\`{e}re and Conjugate Descent variants of the conjugate gradient method is guaranteed due to the boundedness of level sets and Lipschitz continuity of the gradient of the function to be minimized.\cite{SunZhang01}.

\section{Simulation of nonrobust beamforming scenario}\label{sec:cls}
To verify the performance of the algebraic solution in the perfect-CSI case, we modeled a source node with $M=4$ antennas,   $\gamma_1=\gamma_2=10$ and $p_1=p_2=10~W$. The  complex channel gain vectors were chosen randomly so that all components were complex Gaussian random variables with variance 1.  
We took $\sigma_2=\sigma_1$, and  used a grid of value pairs $(\sigma_1,\sigma_R)$ covering the range $1 \le \sigma_1, \sigma_R \le 2$. 5000 simulations were performed for each  $(\sigma_1,\sigma_R)$. For each simulation, the semidefinite-programming solution was computed using the Matlab-based convex optimization system \texttt{cvx} \cite{cvx}, and cases for  which the exact solution required power of more than 25~W were discarded. For the remaining cases, the conjugate gradient algorithm was used to estimate the algebraic solution corresponding to the `$+$' sign in (\ref{eq:UncSoln}). The starting point for the conjugate gradient was chosen as $x=y=0$, which corresponds to the ``maximal-ratio receive, maximal-ratio transmit'' (MRR-MRT)  suboptimal solution in \cite{Rui09}. Iteration was terminated when  the power reduction achieved by the latest iteration was less than 0.5 percent. 

\begin{figure}[ht]
\begin{center}
\includegraphics[width=3.1in]{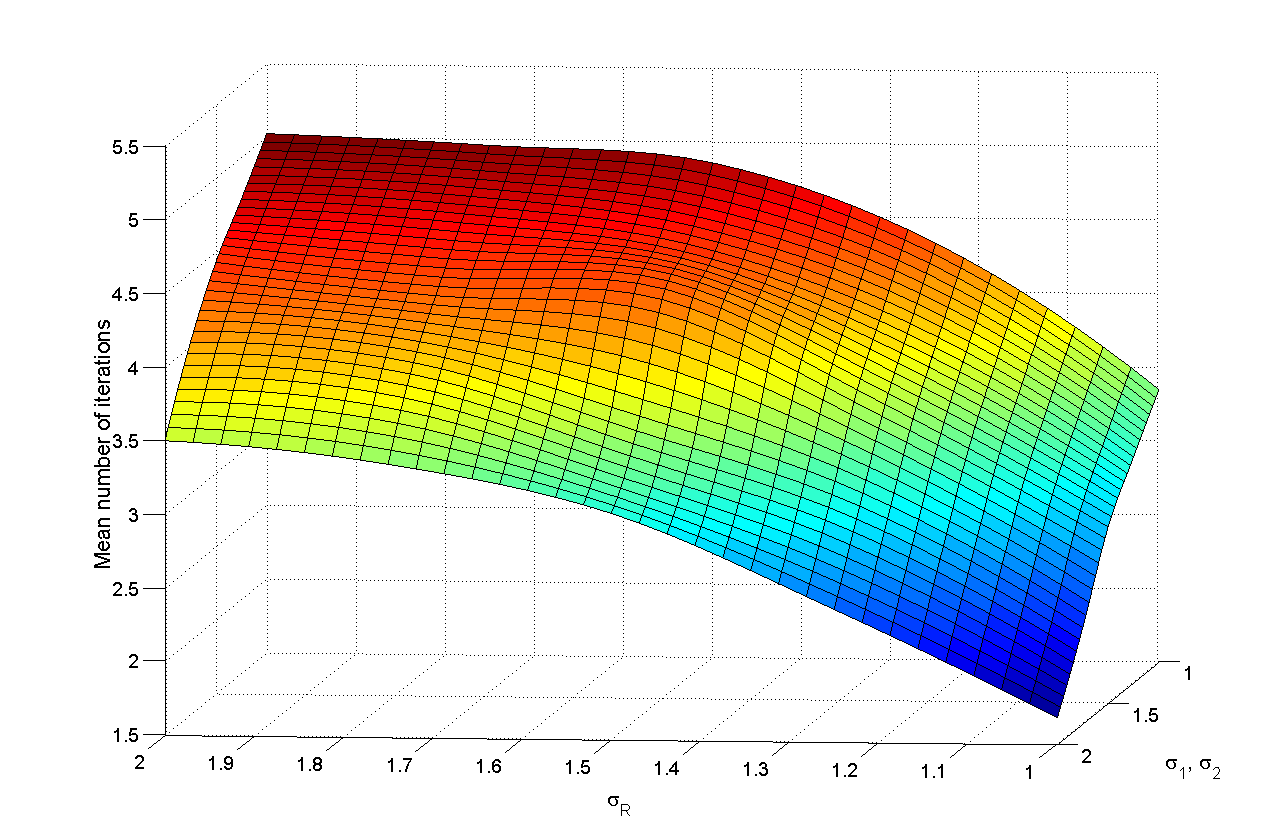}
\caption{Number of iterations until convergence for conjugate-gradient solution.}\label{fig:NumberOfIterations}
\end{center}
\end{figure}

Over the entire range of parameter values, the algebraic solution evaluated using conjugate gradient gave average power increases of less than 0.008~dB over the optimal semidefinite-programming solution.  Convergence of the conjugate gradient solution required 2-5 iterations (on average) over the range of parameter values, as shown in Figure~\ref{fig:NumberOfIterations}. These results confirm that t the solution of  (\ref{eq:UncSoln}) with the `$+$' sign is optimal.  

\begin{figure}[ht]
\begin{center}
\includegraphics[width=4in]{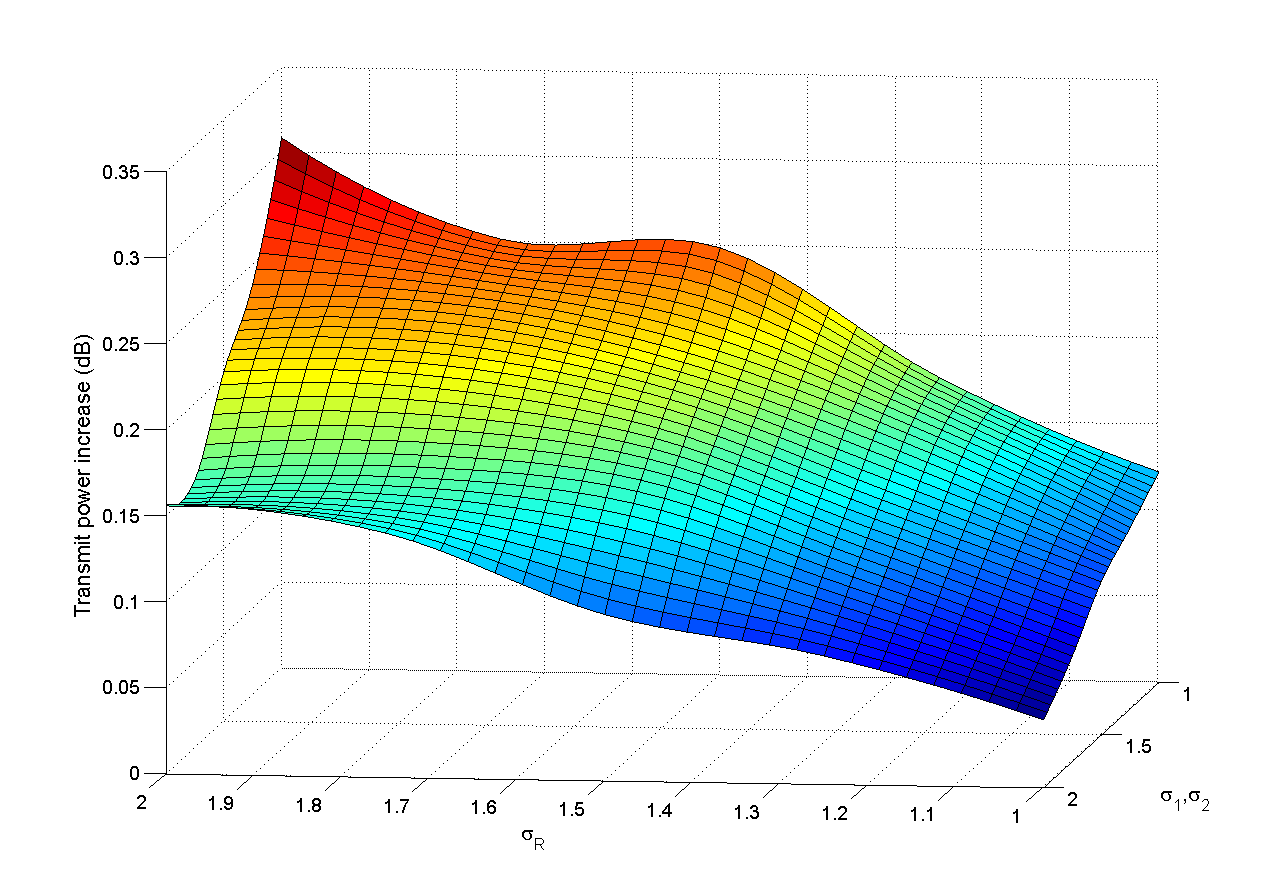}
\caption{Transmit power increase over exact solution from MRR-MRT solution.}\label{fig:performance}
\end{center}
\end{figure}

The suboptimal MRR-MRT solution also performed very well, and produced power increases of only  0.05-0.35~dB as shown in Figure~\ref{fig:performance}.  
Although the improvement of the conjugate gradient solution over the MRR-MRT solution is not great, it comes at very low cost: each conjugate gradient iteration requires only about 300 MAC operations, as compared to about 300 operations for the MRR-MRT solution itself. In contrast, the complexity of an  exact solution via convex programming was computed in \cite{Aziz14} as over 500,000 operations ($\mathcal{O}(k^3+n^3k+n^2k^2)$ with $k=64, n=8$).

%
%

\section{Simulation of robust beamforming scenario}\label{sec:robust}

Reference \cite{Aziz14} demonstrates that given a perfect-CSI solution, a low-complexity suboptimal solution with very high performance can be found for the  robust case with imperfect CSI.  In this section, we compare the performance of suboptimal robust solutions based on each of the three nonrobust solutions modeled in the previous section. 

Simulations were performed for a robust beamforming scenario with $M= 4$, ${\sigma^{2}_{R}}=1 W$, ${\sigma^{2}_{i}}=1 W$, $, p_{i}=10 W$, $\gamma_i=10$,
and $\epsilon_i = [0.01, 0.15]  $ with increments of $0.02$  $(i=1,2)$.
The channel was generated as $\hat{\boldsymbol{h}}_{i}\sim \mathcal{CN}(0, \boldsymbol{I})$; and the channel estimation error $\Delta\boldsymbol{h}_{i}$ was generated as  $\Delta\boldsymbol{h}_{i}\sim \mathcal{CN}(0, (\epsilon_i^2/16)\boldsymbol{I})$, which corresponds to a probability of 0.958 that $\|\Delta \boldsymbol{h}_i \| < \epsilon_i, i=1,2$). 

\begin{figure}[ht]
\begin{center}
\includegraphics[width=2.8in]{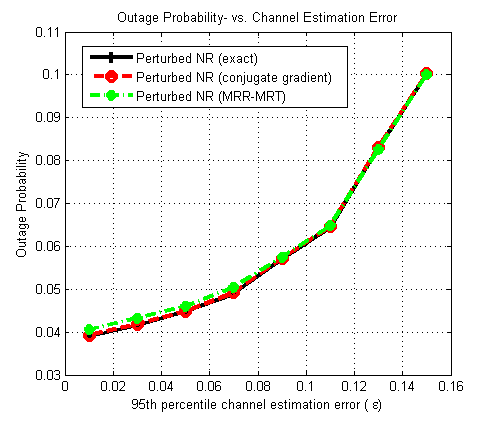}
\caption{TWR beamforming performance: outage vs. channel estimation
error parameter.}\label{fig:Outage}
\end{center}
\end{figure}

\begin{figure}[ht]
\begin{center}
\includegraphics[width=2.8in]{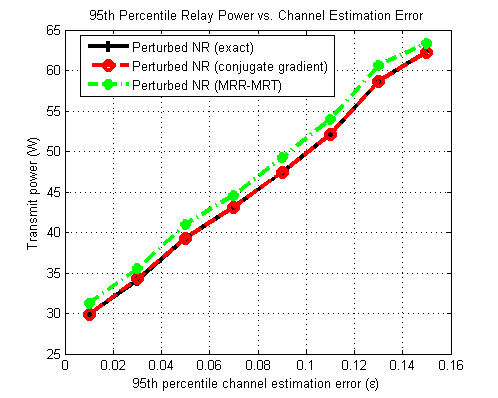}
\caption{TWR beamforming performance: power vs. channel estimation
error parameter.}\label{fig:Power}
\end{center}
\end{figure}

 In the simulations an outage was declared when the SINR at either source node fell below $\gamma_i$. In Fig. (\ref{fig:Outage}), and (\ref{fig:Power}), we respectively plot the outage probability and $95^{th}$ percentile of the empirical cumulative density function (cdf)\cite{ecdf} of the transmit power required to achieve the corresponding outage performance. The perturbed nonrobust solution described in \cite{Aziz14} is used with each of the three nonrobust solutions: the semidefinite programming, algebraic, and MRR-MRT cases are indicated by ``exact'', ``conjugate gradient'', and ``MRR-MRT'' in the figures. As in the previous section, only one of the six candidate algebraic solutions was computed. There was virtually no difference between the performance of the exact and conjugate gradient solutions: this confirms the result of the previous section that one particular candidate of the six candidate solutions almost always gives the best overall solution.  The simulation also shows that  the conjugate gradient solution brings  slight outage reductions (0-3\%) and power reductions (1.5-4.5\%) over the  MRR-MRT approximation.

\section{Conclusions}\label{sec:final}
 
Simulation results show that the conjugate gradient implementation effectively gives the exact solution to the beamforming problem (\ref{modOptProblem0}). Furthermore, it may be used to construct a low-cost, high-performance solution to the corresponding robust problem. In the case of 2-way relays, the algebraic solution provides only limited performance improvements over the even lower-cost MRR-MRT solution. Nonetheless, our results demonstrate mathematical techniques for obtaining computationally inexpensive exact solutions to a nonconvex beamforming-optimization problem.  Similar techniques may be applied to other situations to obtain lower-cost, higher-performing beamforming solutions.
\bibliographystyle{IEEEtran}
\bibliography{ref}

\begin{thebibliography}{10}
\providecommand{\url}[1]{#1}
\csname url@samestyle\endcsname
\providecommand{\newblock}{\relax}
\providecommand{\bibinfo}[2]{#2}
\providecommand{\BIBentrySTDinterwordspacing}{\spaceskip=0pt\relax}
\providecommand{\BIBentryALTinterwordstretchfactor}{4}
\providecommand{\BIBentryALTinterwordspacing}{\spaceskip=\fontdimen2\font plus
\BIBentryALTinterwordstretchfactor\fontdimen3\font minus
  \fontdimen4\font\relax}
\providecommand{\BIBforeignlanguage}[2]{{%
\expandafter\ifx\csname l@#1\endcsname\relax
\typeout{** WARNING: IEEEtran.bst: No hyphenation pattern has been}%
\typeout{** loaded for the language `#1'. Using the pattern for}%
\typeout{** the default language instead.}%
\else
\language=\csname l@#1\endcsname
\fi
#2}}
\providecommand{\BIBdecl}{\relax}
\BIBdecl

\bibitem{Aziz14}
A.~Aziz, C.~Thron, S.~Cui, and C.~Georghiades, ``Linearized robust beamforming
  for two-way relay systems,'' \emph{IEEE Signal Process. Letters}, vol.~21,
  no.~8, pp. 1017 -- 1021, Aug. 2014.

\bibitem{ANC}
S.~Katti, S.~Gollakota, and D.~Katabi, ``Embracing wireless interference:
  Analog networking coding,'' Computer Science and Artificial Intelligence
  Laboratory Technical Report, MIT-CSAIL-TR-2007-012, Feb. 2007.

\bibitem{Rui09}
R.~Zhang, Y.-C. Liang, C.~Choy, and S.~Cui, ``Optimal beamforming for two-way
  multi-antenna relay channel with analogue network coding,'' \emph{IEEE J.
  Sel. Areas Commun.}, vol.~27, no.~5, pp. 699--712, Jun. 2009.

\bibitem{Boyd}
S.~Boyd and L.~Vandenberghe, \emph{Convex Optimization}.\hskip 1em plus 0.5em
  minus 0.4em\relax Cambridge University Press, 2003.

\bibitem{Zeng11}
M.~Zeng, R.~Zhang, and S.~Cui, ``On design of distributed beamforming for
  two-way relay networks,'' \emph{IEEE Trans. Signal Process.}, vol.~59, no.~5,
  pp. 2284--2295, May 2011.

\bibitem{Thron14}
C.~Thron and A.~Aziz, ``Very low complexity algorithms for beamforming in
  two-way relay systems,'' \emph{Imhotep Mathematical Proceedings}, vol.~2,
  no.~1, pp. 13--24, May 2015.

\bibitem{SunZhang01}
J.~Sun and J.~Zhang, ``Global convergence of conjugate gradient methods without
  line search,'' \emph{Annals of Operations Research}, 2001.

\bibitem{cvx}
\BIBentryALTinterwordspacing
I.~CVX~Research, \emph{CVX: Matlab Software for Disciplined Convex
  Programming}, 2015, [Online; accessed 17-August-2015]. [Online]. Available:
  \url{\url{https://http://cvxr.com/cvx/}}
\BIBentrySTDinterwordspacing

\bibitem{AzizICC}
A.~Aziz, M.~Zeng, J.~Zhou, C.~Georghiades, and S.~Cui, ``Robust beamforming
  with channel uncertainty for two-way relay networks,'' in \emph{proceedings
  of IEEE International Conference on Communications (ICC)}, 2012, pp.
  3632--3636.

\bibitem{ecdf}
A.~W. van~der Vaart, \emph{Asymptotic Statistics}.\hskip 1em plus 0.5em minus
  0.4em\relax Cambridge University Press, September 2000.

\end{thebibliography}
\end{document}